\documentclass[showpacs,preprintnumbers,amsmath,amssymb]{revtex4}
\usepackage{amsmath,amssymb,graphics,epsfig,subfigure}
\usepackage{color}
\usepackage{ulem}

\begin{document}
\renewcommand{\baselinestretch}{1.3}

\title{Universal thermodynamic relations with constant corrections for rotating AdS black holes}

\author{Shao-Wen Wei$^1$ \footnote{weishw@lzu.edu.cn}}
\author{Ke Yang$^2$ \footnote{keyang@swu.edu.cn}}
\author{Yu-Xiao Liu$^1$ \footnote{liuyx@lzu.edu.cn}}

\affiliation{$^1$ Institute of Theoretical Physics $\&$ Research Center of Gravitation, Lanzhou University, Lanzhou 730000, People's Republic of China\\
$^2$ School of Physical Science and Technology, Southwest University, Chongqing 400715, China}

\begin{abstract}
In [Phys. Rev. Lett. 124, 101103 (2020)], a universal relation between corrections to entropy and extremality was proposed. The relation was also found to exactly hold for the four-dimensional charged AdS black hole. In this paper, we extend the study to the rotating BTZ and Kerr-AdS black holes when a constant correction to General Relativity is considered for the first time. The entropy and extremality bound are calculated, and they have a closely dependent behavior with the coupling parameter of the constant correction. We confirm the universal relation for the rotating AdS black holes. Furthermore, taking into consideration of the shift of the angular momentum, we confirm one more new universal relation for the rotating cases. In particular, we state a conjecture on a universal relation, which gives a universal conjecture relation between the shifted thermodynamic quantities for arbitrary black hole background. We believe that these universal relations will shed new light on the region of the quantum gravity.
\end{abstract}


\pacs{...}

\maketitle

\section{Introduction}

The ultimate aim of swampland program \cite{Vafa} is to systematically delineate regions and to understand what subset of the infinite space in effective field theory is consistent with quantum gravitational ultraviolet completion. Among the criteria, Weak Gravity Conjecture (WGC) has recently attracted a considerable interest \cite{Hamed}.

The WGC reveals that there exist the states that their mass $M$ is smaller than their charge $Q$, i.e., $M<Q$, which means that ``gravity is the weakest force" \cite{Hamed}. To understand it, one fruitful bed is black hole, and its thermodynamics is generally believed to be in the semiclassical regime. However, for a charged black hole, the mass and charge obey the relation
\begin{equation}
 \frac{Q}{M}\leq 1, \label{cm}
\end{equation}
with the extremal black hole saturating the bound. The naked singularity will disobey this equation and is consistent with that of WGC. However it is forbidden by the cosmic censorship. A possible solution to this problem is to include some correction terms in the action, which will modify the black hole solution and possibly inverse the charge to mass ratio (\ref{cm}). This goal was achieved in Ref. \cite{Kats}, where higher-derivative corrections were included. The result shows that the charge to mass ratio could be larger than one with a combination of coefficients of the higher-derivative operators. The idea that considers the higher-derivative corrections is now known as the ``Black Hole Weak Gravity Conjecture''. Some works \cite{Cheung,Gary,Clifford,Junghans,Hamada,Reall,Bellazzini,Cheungb,Aalsma,Charles,Callum,Gregory,Loges} have been done on this idea to bound the effective field theory coefficients under the WGC.

Another novel proof of the WGC in flat space was given in \cite{Clifford} by considering the shift of the Wald entropy. The authors showed that for a near-extremal black hole case, the shift to the extremality bound at fixed charge and temperature is proportional to the shift in entropy at fixed charge and mass. Then they concluded that the higher-derivative corrections will increase the entropy, and thus the black hole WGC was proved.

In particular, Goon and Penco proposed a universal thermodynamic extremality relation \cite{Goon}
\begin{equation}
 \frac{\partial M_{ext}(\vec{\mathcal{Q}},\epsilon)}{\partial \epsilon}=\lim_{M\rightarrow M_{ext}}
 -T\left(\frac{\partial S(M,\vec{\mathcal{Q}},\epsilon)}{\partial \epsilon}\right)_{M,\vec{\mathcal{Q}}},\label{GPr}
\end{equation}
where $\epsilon$ is the control parameter of the correction, $M_{ext}$, $T$, and $S$ are, respectively, the mass bound, temperature, and entropy of the black hole after the correction, and $\vec{\mathcal{Q}}$ denotes the extensive quantities of the black hole thermodynamics. For instance, $\vec{\mathcal{Q}}$ is the charge and angular momentum for the charged and rotating black holes, respectively. It was also found that this relation is exactly hold for the four-dimensional charged AdS black hole. This relation displays the black hole WGC-like behavior. Subsequently, the authors of Ref.~\cite{Cremonini} computed the four-derivative corrections to the geometry of charged AdS black holes for general dimensions and horizon geometries, and confirmed the Goon-Penco relation (\ref{GPr}). Other related works can be found in Refs. \cite{Cano,Chimento}.

So far, however, these ideas have not fully made their way to the rotating black holes in AdS space. Therefore, in this paper, we will focus on the Goon-Penco relation (\ref{GPr}) in the background of rotating BTZ and Kerr-AdS black holes when a constant correction, which can also be understood as a leading-order correction, is added to the black hole action. Furthermore, the rotating black holes also provide us a good chance to examine whether some new equality holds when the angular momentum shifts. Therefore, we will also check the following new angular momentum-extremality relation as an attempt,
\begin{equation}
 \left(\frac{\partial M_{ext}}{\partial \epsilon}\right)_{J,l}=\lim_{M\rightarrow M_{ext}}
 -\Omega\left(\frac{\partial J}{\partial \epsilon}\right)_{M,S,l}.\label{GPr2}
\end{equation}

The paper is organized as follows. In Sec. \ref{iner}, we plan to check the extremality relations (\ref{GPr}) and (\ref{GPr2}) for the rotating BTZ black hole. Then we apply it to the rotating Kerr-AdS black hole in Sec. \ref{iners}. Finally, we end up this paper with a brief summary and give a conjecture of an extended extremality relation.

\section{Rotating BTZ black hole}
\label{iner}

Here we consider the three-dimensional rotating BTZ black hole \cite{BTZ}. The corresponding action reads
\begin{equation}
 S=\frac{1}{16\pi G}\int \sqrt{-g}d^3x(R-2\Lambda).
\end{equation}
Here $G$ is the the gravitational constant in three-dimensional spacetime and will be set to $G=1/8$ for simplicity, and the cosmological constant $\Lambda$ relates to the AdS radius $l$ as $\Lambda=-1/l^{2}$. Solving the Einstein field equation, the line element for the black hole can be obtained as
\begin{equation}
 ds^{2}=-f(r)dt^{2}+\frac{1}{f(r)}dr^{2}+r^{2}\left(d\phi-\frac{J}{2r^{2}}dt\right)^{2},
\end{equation}
where the metric function is given by
\begin{equation}
 f(r)=-M+\frac{J^2}{4r^2}+\frac{r^2}{l^2}.
\end{equation}
The parameters $M$ and $J$ are the mass and angular momentum, respectively. The outer and inner horizons are located at $f(r_{\pm})=0$, where
\begin{equation}
 r_{\pm}=\sqrt{\frac{Ml^2\pm\sqrt{M^2l^4-J^2l^2}}{2}}.
\end{equation}
The black hole entropy $S=4\pi r_{+}$. Further, the black hole mass, temperature, and angular velocity can be expressed in terms of $S$ and $J$ as
\begin{eqnarray}
 M&=&\frac{4 \pi ^2 J^2}{S^2}+\frac{S^2}{16 \pi^2 l^2},\\
 T&=&\frac{S}{8 \pi ^2 l^2}-\frac{8 \pi ^2 J^2}{S^3},\\
 \Omega&=&\frac{8\pi^2 J}{S^2}.
\end{eqnarray}
Now we consider a small constant correction to the action,
\begin{equation}
 \Delta S=-\frac{1}{2\pi}\int\sqrt{-g} d^3x(\epsilon*Const).
\end{equation}
Here we assume that $\epsilon$ is a small  parameter, and when $\epsilon\rightarrow0$, we get the uncorrected action. For simplicity, we can reformulate the action such that
\begin{equation}
 \Delta S=-\frac{1}{2\pi}\int\sqrt{-g} d^3x(2\epsilon\Lambda).
\end{equation}
Due to the correction, the black hole solution will be shifted. The shifted mass and  temperature will be of the following forms
\begin{eqnarray}
 M&=&\frac{4 \pi ^2 J^2}{S^2}+\frac{S^2(1+\epsilon)}{16 \pi^2 l^2}, \label{mas}\\
 T&=&\frac{S(1+\epsilon)}{8 \pi ^2 l^2}-\frac{8 \pi ^2 J^2}{S^3}.\label{tem}
\end{eqnarray}
The extremality bound for this black hole will also be modified. By solving $T=0$, the new extremality can be obtained as
\begin{equation}
 S=\frac{2\pi \sqrt{2Jl}}{\sqrt[4]{\epsilon +1}}.\label{extremality}
\end{equation}
It is clear that the extremal entropy bound is decreased by a positive correction parameter $\epsilon$. In the following, we aim to check the relation (\ref{GPr}). Solving $\epsilon$ from (\ref{mas}), one gets
\begin{equation}
 \epsilon=\frac{16 l^2 \left(\pi ^2 M S^2-4 \pi ^4 J^2\right)}{S^4}-1.\label{epsi}
\end{equation}
Taking the derivative of it with respect to $S$, one can get
\begin{equation}
 \left(\frac{\partial\epsilon}{\partial S}\right)_{M,J,l}=\frac{1}{S^5} \left(128\pi ^4 J^2 l^2-2 S^4 (\epsilon +1)\right). \label{es}
\end{equation}
Combining (\ref{tem}) and (\ref{es}), we obtain
\begin{equation}
 -T\left(\frac{\partial S}{\partial \epsilon}\right)_{M,J,l}=\frac{S^2}{16\pi^2 l^2}. \label{relation1}
\end{equation}
By making use of (\ref{extremality}) at the extremality, the above relation (\ref{relation1}) is reduced to
\begin{equation}
 \lim_{M\rightarrow M_{ext}} -T\left(\frac{\partial S}{\partial \epsilon}\right)_{M,J,l}=\frac{J}{2 l \sqrt{\epsilon+1}}.\label{ms}
\end{equation}
In order to calculate the left side of (\ref{GPr}), we insert the entropy at the extremality into the mass (\ref{mas}) to obtain the mass bound,
\begin{equation}
 M_{ext}=\frac{J \sqrt{\epsilon +1}}{l}.\label{mex}
\end{equation}
It is clear that a positive constant correction parameter $\epsilon$ will increase the mass bound, so this is a hint that the correction could satisfy the corresponding WGC condition for the rotating black hole. After the differentiation, we have
\begin{equation}
 \left(\frac{\partial M_{ext}}{\partial \epsilon}\right)_{J,l}=\frac{J}{2 l \sqrt{\epsilon+1}},
\end{equation}
which is exactly the same with (\ref{ms}). Therefore, we confirm the Goon-Penco extremality relation (\ref{GPr}) for the rotating BTZ black hole under the constant correction of the action:
\begin{equation}
 \left(\frac{\partial M_{ext}}{\partial \epsilon}\right)_{J,l}=\lim_{M\rightarrow M_{ext}}
 -T\left(\frac{\partial S(M,J,\epsilon)}{\partial \epsilon}\right)_{M,J,l}.
\end{equation}
Next, we would like to examine the second relation (\ref{GPr2}). By using (\ref{extremality}) and (\ref{mex}), we can obtain the shift of the mass bound
\begin{equation}
 \left(\frac{\partial M_{ext}}{\partial \epsilon}\right)_{J,l}=\frac{S^2}{16 \pi^2 l^2}.\label{btz}
\end{equation}
With the help of (\ref{epsi}), it is easy to get
\begin{equation}
 \left(\frac{\partial J}{\partial \epsilon}\right)_{M,S,l}=-\frac{S^4}{128 \pi ^4 J l^2}.
\end{equation}
Multiplying by the angular velocity and taking the extremality, we have
\begin{equation}
 \lim_{M\rightarrow M_{ext}} -\Omega\left(\frac{\partial J}{\partial \epsilon}\right)_{M,S,l}=\frac{S^2}{16 \pi^2 l^2}.
\end{equation}
Comparing with (\ref{btz}), it is obvious that our angular momentum-extremality relation (\ref{GPr2}) is exactly confirmed.

\section{Rotating Kerr-AdS black hole}
\label{iners}

Now, we would like to consider the rotating Kerr-AdS black hole case in four dimensions following the same process showed above.

\begin{figure}
\center{
\includegraphics[width=7cm]{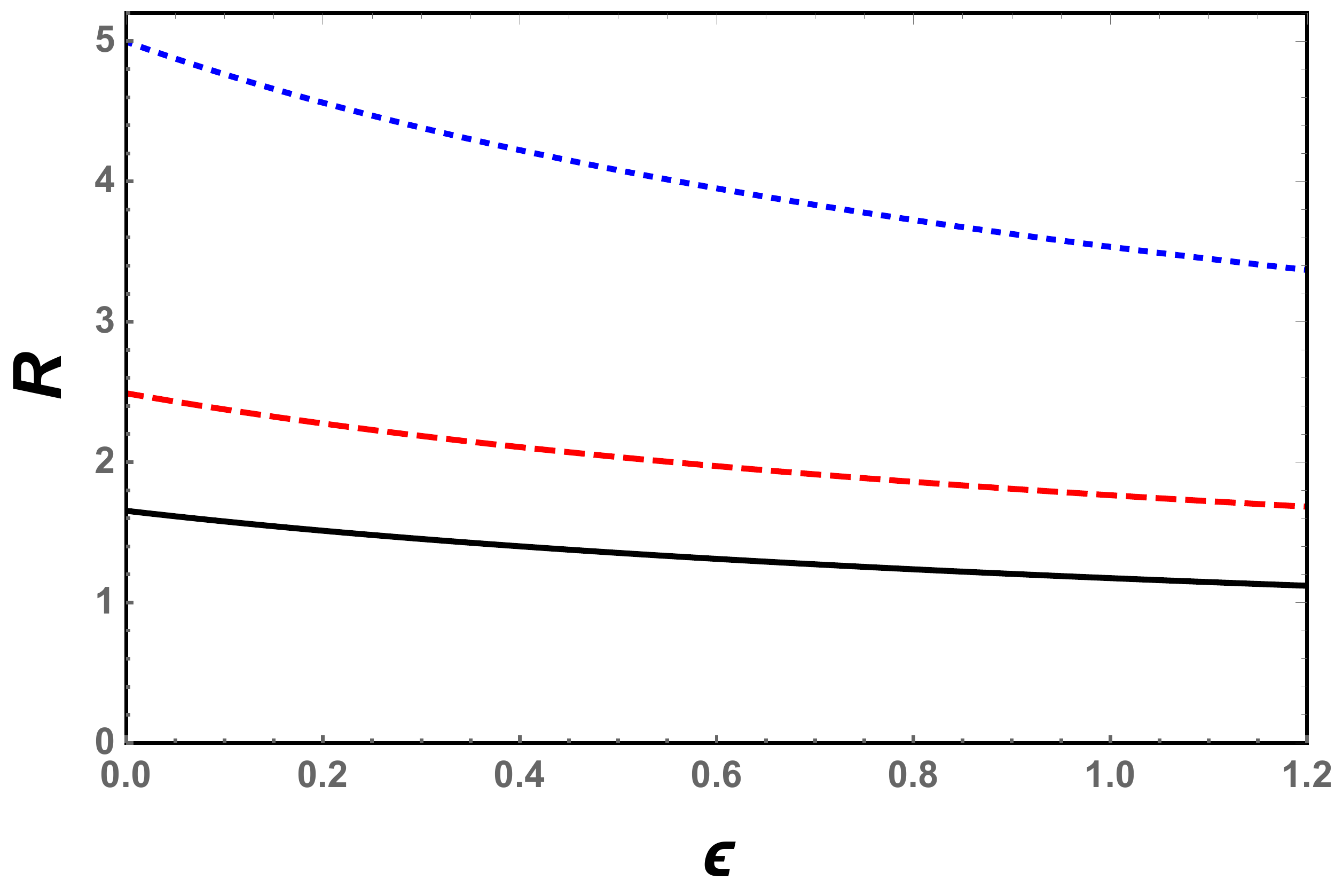}}
\caption{Behaviors of $R\equiv\lim_{M\rightarrow M_{\text{ext}}}-T\left(\frac{\partial S}{\partial \epsilon}\right)_{M}$ (lines) with fixed angular momentum $J=1$, and $l$=0.1, 0.2, and 0.3 from top to bottom. }\label{ppPT}
\end{figure}

The line element of the Kerr-AdS black hole is
\begin{eqnarray}
 ds^{2}=-\frac{\Delta}{\rho^{2}}\bigg(dt-\frac{a\sin^{2}\theta}{\Xi}d\varphi\bigg)^{2}
        +\frac{\rho^{2}}{\Delta}dr^{2}+\frac{\rho^{2}}{1-a^{2}/l^{2}\cos^{2}\theta}d\theta^{2}\nonumber\\
        +\frac{(1-a^{2}/l^{2}\cos^{2}\theta)\sin^{2}\theta}{\rho^{2}}\bigg(adt-\frac{r^{2}+a^{2}}{\Xi}d\varphi\bigg)^{2},
        \label{Kerr-AdSHBsolution}
\end{eqnarray}
with the metric functions given by
\begin{eqnarray}
 \rho^{2}&=&r^{2}+a^{2}\cos^{2}\theta,\quad
 \Xi=1-\frac{a^{2}}{l^{2}},  \label{Kerr-AdSHBsolutionrho2} \\
 \Delta&=&(r^{2}+a^{2})(1+r^{2}/l^{2})-2Mr.\label{Kerr-AdSHBsolutionDelta}
\end{eqnarray}
This solution is obtained by solving the following action 
\begin{equation}
 S=\frac{1}{16\pi}\int \sqrt{-g}d^4x\left(R+\frac{6}{l^2}\right).
\end{equation}
Note that, different form the statement above, $G$ is set to be 1 in this section.
Now we consider a small constant correction and reformulate the action as
\begin{equation}
 S=\frac{1}{16\pi}\int \sqrt{-g}d^4x\left(R+(1+\epsilon)\frac{6}{l^2}\right),
\end{equation}
with $\epsilon$ the small correction parameter. The Kerr-AdS black hole solution (\ref{Kerr-AdSHBsolution}-\ref{Kerr-AdSHBsolutionDelta}) will be recovered when $\epsilon=0$. For a nonvanishing $\epsilon$, the shifted mass can be obtained according to the Christodoulou-Ruffini-like squared-mass formula for the Kerr-AdS black hole \cite{Caldarelli}, namely,
\begin{equation}
 M=\sqrt{\frac{\left(\pi  l^2+S \epsilon +S\right)
   \left(4 \pi ^3 J^2 l^2+S^2 \left(\pi l^2+S \epsilon +S\right)\right)}{4 \pi ^3 l^4 S}}. \label{masss}
\end{equation}
Employing the first law, the temperature and  angular velocity can be calculated as
\begin{eqnarray}
 T&=&\frac{S^2 \left(\pi  l^2+S \epsilon +S\right)\left(\pi  l^2+3 S (\epsilon +1)\right)-4
   \pi ^4 J^2 l^4}{4 \pi ^{3/2} l^2 \sqrt{S^3\left(\pi  l^2+S \epsilon +S\right)
   \left(4 \pi ^3 J^2 l^2+S^2 \left(\pi l^2+S \epsilon +S\right)\right)}},\label{temm}\\
 \Omega&=&2 \pi ^{3/2} J \sqrt{\frac{\pi  l^2+S
   \epsilon +S}{4 \pi ^3 J^2 l^2 S+S^3
   \left(\pi  l^2+S \epsilon +S\right)}}.
\end{eqnarray}
Solving the mass equation (\ref{masss}), the parameter $\epsilon$ is 
\begin{equation}
 \epsilon=-\frac{\pi  l^2 \left(-2 \sqrt{\pi^4 J^4+\pi M^2 S^3}+2 \pi ^2 J^2+S^2\right)}{S^3}-1.\label{ese}
\end{equation}
After a simple calculation, we have
\begin{equation}
 \left(\frac{\partial\epsilon}{\partial S}\right)_{M,J,l}=\frac{4 \pi ^4 J^2 l^4-S^2 \left(\pi  l^2+S
   \epsilon +S\right) \left(\pi  l^2+3 S(\epsilon +1)\right)}{4 \pi ^3 J^2 l^2 S^2+2 S^4 \left(\pi  l^2+S \epsilon+S\right)}.\label{ess}
\end{equation}
Combining with (\ref{temm}) and (\ref{ess}), we obtain
\begin{equation}
 -T\left(\frac{\partial S}{\partial \epsilon}\right)_{M,J,l}=\frac{\sqrt{S}\left(2 \pi ^3 J^2 l^2+S^2 \left(\pi l^2+S \epsilon +S\right)\right)}{2 \pi ^{3/2} l^2
   \sqrt{\left(\pi  l^2+S \epsilon
   +S\right) \left(4 \pi ^3 J^2 l^2+S^2
   \left(\pi  l^2+S \epsilon
   +S\right)\right)}}.\label{tsss}
\end{equation}
Now, we turn to consider the extremality for the black hole solution. An extremal black hole has a vanishing temperature, and thus these quantities satisfy the following equation
\begin{equation}
 3 (\epsilon +1)^2S^4 +4 \pi  l^2 (\epsilon +1)S^3 +\pi ^2 l^4 S^2-4 \pi ^4 J^2 l^4=0.\label{seeq}
\end{equation}
Obviously, this equation admits four roots of $S$, which are
\begin{eqnarray}
 S_{1,2}=-\frac{\pi l^2}{3(1+\epsilon)}-Q\pm\frac{1}{2}\sqrt{-4Q^2+\frac{2\pi^2l^4}{3(1+\epsilon)^2}+\frac{2 \pi ^3 l^6}{27 (\epsilon +1)^3Q}},\\
 S_{3,4}=-\frac{\pi l^2}{3(1+\epsilon)}-Q\pm\frac{1}{2}\sqrt{-4Q^2+\frac{2\pi^2l^4}{3(1+\epsilon)^2}-\frac{2 \pi ^3 l^6}{27 (\epsilon +1)^3Q}},
\end{eqnarray}
where
\begin{eqnarray}
 Q&=&\frac{ \pi  l^2}{6(1+\epsilon)}\sqrt{\frac{\left(\sqrt[3]{P-432 J^2 L^2
   (\epsilon +1)^2+l^6}+L^2\right)^2-144 J^2
   (\epsilon +1)^2}{\sqrt[3]{P l^6-432 J^2
   l^8 (\epsilon +1)^2+l^{12}}}},\\
 P&=&12 \sqrt{3} J (\epsilon +1) \sqrt{6912 J^4
   (\epsilon +1)^4+288 J^2 l^4 (\epsilon
   +1)^2-l^8}.
\end{eqnarray}
A detailed analysis shows that $S_3$ and $S_4$ are imaginary, and $S_2$ is negative. So $S_1$ denotes the extremality. Inserting $S_1$ into (\ref{tsss}), we can obtain the right side of (\ref{GPr}) for the Kerr-AdS black hole, i.e., $R\equiv\lim_{M\rightarrow M_{\text{ext}}}-T\left(\frac{\partial S}{\partial \epsilon}\right)_{M}$. However, instead of  listing the cumbersome calculations, here we illustrate them in Fig. \ref{ppPT},  where the angular momentum $J=1$, and $l$=0.1, 0.2, and 0.3 from top to bottom. As shown in the figure, we find that $R$ slowly decreases with $\epsilon$ for fixed $J$ and $l$, and moreover, $R$ also decreases with $l$.

\begin{figure}
\center{
\includegraphics[width=7cm]{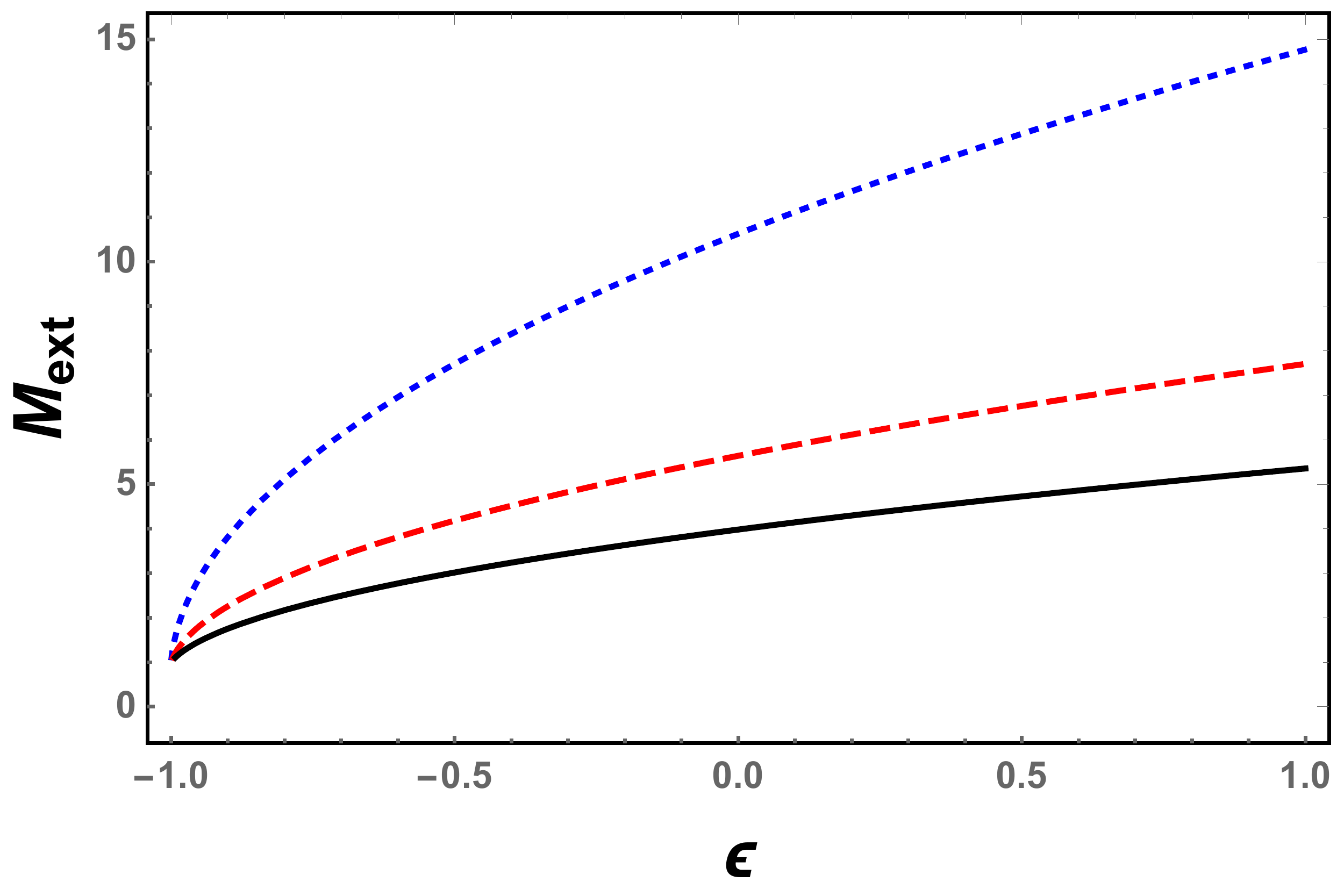}}
\caption{Mass bound for the rotating Kerr-AdS black hole with a fixed angular momentum $J=1$. From top to bottom, the AdS radius $l$=0.1, 0.2, and 0.3, respectively.}\label{ppPTphase}
\end{figure}

Plugging $S_{1}$ into (\ref{masss}), we get the mass bound for the Kerr-AdS black hole. The result is plotted in Fig. \ref{ppPTphase}. For a fixed $J$, the figure shows that the mass bound increases with the parameter $\epsilon$. This behavior is the same as that of the rotating BTZ black hole. It is also straightforward to calculate the left side of (\ref{GPr}) at the extremality for the Kerr-AdS black hole, which is given by
\begin{equation}
 \left(\frac{\partial M_{ext}}{\partial \epsilon}\right)_{J,l}
 =\frac{4 \pi ^3 J^2 l^2 S_1^2
        +(\partial_{\epsilon}S_{1})
       \left[S_1^2 \left(\pi  l^2+S_1 (\epsilon   +1)\right) 
      \left(\pi  l^2+3 S_1 (\epsilon    +1)\right)
            -4 \pi ^4 J^2 l^4\right]
      +2 \pi  l^2 S_1^4+2 S_1^5 (\epsilon +1)}
   {4 \pi^{3/2} l^2 S_1^{3/2} 
     \sqrt{\left[\pi l^2+S_1 (\epsilon +1)\right] 
           \left[4 \pi^3 J^2 l^2+\pi  l^2 S_1^2+S_1^3 (\epsilon +1)\right]}
   }.\label{mmm}
\end{equation}
Considering that $S_1$ satisfies (\ref{seeq}), the above equation will be nicely reduced to
\begin{equation}
 \left(\frac{\partial M_{ext}}{\partial \epsilon}\right)_{J,l}
 =\frac{\sqrt{S_1}\left(2 \pi ^3 J^2 l^2+S_1^2 \left(\pi l^2+S_1 \epsilon +S_1\right)\right)}{2 \pi ^{3/2} l^2
   \sqrt{\left(\pi  l^2+S_1 \epsilon
   +S_1\right) \left(4 \pi ^3 J^2 l^2+S_1^2
   \left(\pi  l^2+S_1 \epsilon
   +S_1\right)\right)}}.\label{was}
\end{equation}
Obviously, comparing with (\ref{tsss}), it equals to $R$. Thus the left and right sides of (\ref{GPr}) meet each other for the Kerr-AdS black hole, and equation (\ref{GPr}) is confirmed.

Next, we continue to check the angular momentum-extremality relation (\ref{GPr2}). By solving $J$ from the vanishing temperature, we obtain another extremality condition
\begin{equation}
 J=\frac{S \sqrt{\pi ^2 l^4+4 \pi  l^2 S
   (\epsilon +1)+3 S^2 (\epsilon +1)^2}}{2
   \pi^2 l^2}.
\end{equation}
Using (\ref{ese}), one gets
\begin{equation}
 \left(\frac{\partial J}{\partial \epsilon}\right)_{M,S,l}=-\frac{2 \pi ^3 J^2 l^2 S+\pi  l^2 S^3+S^4
   \epsilon+S^4}{4 \pi ^3 J l^2 \left(\pi
    l^2+S \epsilon +S\right)}.
\end{equation}
Further, we can obtain the following result at the extremality
\begin{equation}
 \lim_{M\rightarrow M_{ext}} -\Omega\left(\frac{\partial J}{\partial \epsilon}\right)_{M,S,l}=\frac{3 S_1^{\frac{3}{2}} \left(\pi  l^2+S_1 \epsilon
   +S_1\right)}{4 \pi ^2 l^3 \sqrt{2 \pi  l^2+3
   S_1 (\epsilon +1)}}.\label{uye}
\end{equation}
In order to check the second extremality relation (\ref{GPr2}). We solve the vanishing temperature in (\ref{temm}), and obtain the following relation
\begin{equation}
 \pi l^2+S\epsilon+S=\frac{4\pi^2J^2l^4}{S^2(\pi l^2+3S\epsilon+3S)}.
\end{equation}
Inserting this relation into (\ref{was}), we find $(\partial_\epsilon M_{ext})_{J,l}$ exactly equals to (\ref{uye}). Thus, we confirm the second extremality relation (\ref{GPr2}) for the rotating Kerr-AdS black hole.

\section{Summary}
\label{Summary}

In this paper, we investigated the thermodynamic relation between the shifted entropy and mass bound when a constant correction to the action is included in the rotating black hole backgrounds.

For both the rotating BTZ and Kerr-AdS black holes, we found that the shifted mass bounds increase with the constant correction parameter $\epsilon$, which indicates that the correction could help the black hole to satisfy the WGC condition as the charged black holes do.

For the first time, we also confirmed the Goon-Penco relation (\ref{GPr}), a universal relation between the shifted entropy and mass bound, in the rotating black hole backgrounds. In particular, this relation holds for three- and four-dimensional spacetime. Furthermore, we proposed a new angular momentum-extremality relation (\ref{GPr2}) for the rotating black holes, and confirmed it for both the rotating BTZ and Kerr-AdS black holes.

As an extension, we would like to suggest the following conjecture. If the first law of the black hole thermodynamics is
\begin{equation}
 dM=TdS+\sum_{i}Y_i dX^i,
\end{equation}
then we conjecture that even the higher-derivative corrections are included in the action, the following extremality relations
\begin{equation}
 \left(\frac{\partial M_{ext}}{\partial \epsilon}\right)_{\sum_{i}X^{i}}=\lim_{M\rightarrow M_{ext}}
 -T\left(\frac{\partial S}{\partial \epsilon}\right)_{M,\sum_{i}X^{i}}
 =\lim_{M\rightarrow M_{ext}}
 -Y_j\left(\frac{\partial X^j}{\partial \epsilon}\right)_{M,S,\sum_{i(i\neq j)}X^{i}},
 \label{Conjecture}
\end{equation}
hold for any black hole solution in any dimensional spacetime when the third law of thermodynamics satisfies. By taking $X^i=J$, it leads to the extremality relation (\ref{GPr2}). We believe that these new relations will shed new light on the region of quantum gravity.

\section*{Acknowledgements}
This work was supported by the National Natural Science Foundation of China (Grants No. 11675064 and No. 11875151). K. Yang acknowledges the support of ``Fundamental Research Funds for the Central Universities" under Grant No. XDJK2019C051.


\begin{thebibliography}{99}

\bibitem{Vafa}
 C. Vafa,
{\em The String landscape and the swampland},
 [arXiv:hep-th/0509212 [hep-th]].

\bibitem{Hamed}
 N. Arkani-Hamed, L. Motl, A. Nicolis, and C. Vafa,
{\em The String landscape, black holes and gravity as the weakest force},
 J. High Energy Phys. \textbf{06}, 060 (2007),
 [arXiv:hep-th/0601001 [hep-th]].

\bibitem{Kats}
 Y. Kats and P. Petrov,
{\em Effect of curvature squared corrections in AdS on the viscosity of the dual gauge theory},
 J. High Energy Phys. \textbf{01}, 044 (2009),
 [arXiv:0712.0743 [hep-th]].

\bibitem{Cheung}
 C. Cheung and G. N. Remmen,
 {\em Naturalness and the Weak Gravity Conjecture},
  Phys. Rev. Lett. \textbf{113}, 051601 (2014),
   [arXiv:1402.2287 [hep-ph]].

\bibitem{Gary}
 G. Shiu, P. Soler, and W. Cottrell,
 {\em Weak Gravity Conjecture and extremal black holes},
 Sci. China Phys. Mech. Astron. \textbf{62}, 110412 (2019),
  [arXiv:1611.06270 [hep-th]].

\bibitem{Clifford}
 C. Cheung, J. Liu, and G. N. Remmen,
 {\em Proof of the Weak Gravity Conjecture from Black Hole Entropy},
  J. High Energy Phys. \textbf{10}, 004 (2018),
  [arXiv:1801.08546 [hep-th]].

\bibitem{Junghans}
 S. Andriolo, D. Junghans, T. Noumi, and G. Shiu,
 {\em A Tower Weak Gravity Conjecture from Infrared Consistency},
  Fortsch. Phys. \textbf{66}, 1800020 (2018),
    [arXiv:1802.04287 [hep-th]].

\bibitem{Hamada}
Y. Hamada, T. Noumi, and G. Shiu,
  {\em Weak Gravity Conjecture from Unitarity and Causality},
  Phys. Rev. Lett. \textbf{123}, 051601 (2019),
   [arXiv:1810.03637 [hep-th]].

\bibitem{Reall}
 H. S. Reall and J. E. Santos,
 {\em Higher derivative corrections to Kerr black hole thermodynamics},
  JHEP \textbf{1904}, 021 (2019),
 [arXiv:1901.11535 [hep-th]].

\bibitem{Bellazzini}
 B. Bellazzini, M. Lewandowski, and J. Serra,
 {\em Amplitudes' Positivity, Weak Gravity Conjecture, and Modified Gravity},
  Phys. Rev. Lett. \textbf{123}, 251103 (2019),
 [arXiv:1902.03250 [hep-th]].

\bibitem{Cheungb}
 C. Cheung, J. Liu, and G. N. Remmen,
 {\em Entropy Bounds on Effective Field Theory from Rotating Dyonic Black Holes},
  Phys. Rev. D \textbf{100}, 046003 (2019),
 [arXiv:1903.09156 [hep-th]].

\bibitem{Aalsma}
 L. Aalsma, A. Cole, and G. Shiu,
 {\em Weak Gravity Conjecture, Black Hole Entropy, and Modular Invariance},
  JHEP \textbf{08}, 022 (2019),
 [arXiv:1905.06956 [hep-th]].


 \bibitem{Charles}
A. M. Charles,
  {\em The Weak Gravity Conjecture, RG Flows, and Supersymmetry},
  Phys. Rev. Lett. \textbf{123}, 051601 (2019),
   [arXiv:1906.07734 [hep-th]].

\bibitem{Callum}
 C. R. T. Jones and B. McPeak,
 {\em The Black Hole Weak Gravity Conjecture with Multiple Charges},
 [arXiv:1908.10452 [hep-th]].

\bibitem{Gregory}
 G. J. Loges, T. Noumi, and G. Shiu,
 {\em Thermodynamics of 4D Dilatonic Black Holes and the Weak Gravity Conjecture},
 [arXiv:1909.01352 [hep-th]].

\bibitem{Loges}
 G. J. Loges, T. Noumi, and G. Shiu,
 {\em Festina Lente: EFT Constraints from Charged Black Hole Evaporation in de Sitter},
 J. High Energy Phys. \textbf{2001}, 039 (2020),
 [arXiv:1910.01648 [hep-th]].

\bibitem{Goon}
G. Goon and R. Penco,
 {\em A Universal Relation Between Corrections to Entropy and Extremality},
 [arXiv:1909.05254 [hep-th]].

\bibitem{Cremonini}
  S. Cremonini, C. R.T. Jones, J. T. Liu, and B. McPeak,
  {\em Higher-Derivative Corrections to Entropy and the Weak Gravity Conjecture in Anti-de Sitter Space},
  [arXiv:1912.11161 [hep-th]].

\bibitem{Cano}
P. A. Cano, T. Ortin, and P. F. Ramirez,
 {\em On the extremality bound of stringy black holes},
   [arXiv:1909.08530 [hep-th]].

\bibitem{Chimento}
P. A. Cano, S. Chimento, R. Linares, T. Ortin, and P. F. Ramirez,
 {\em $\alpha'$ corrections of Reissner-Nordstr\"om black holes},
    J. High Energy Phys. \textbf{2002}, 031 (2020) 031,
   [arXiv:1910.14324 [hep-th]]. 

\bibitem{BTZ}
 M. Banados, C. Teitelboim, and J. Zanelli,
 {\em The Black hole in three-dimensional space-time},
    Phys. Rev. Lett. \textbf{69}, 1849 (1992) 1849,
   [arXiv:hep-th/9204099].

\bibitem{Caldarelli}
 M. M. Caldarelli, G. Cognola, and D. Klemm,
 {\em Thermodynamics of Kerr-Newman-AdS black holes and conformal field theories},
    Class. Quant. Grav. 17, 399 (2000),
    [arXiv:hep-th/9908022].


\end{thebibliography}
\end{document}